\documentstyle[twoside, epsf]{article}

\input ibvs2.sty
\input ibvst.sty
\input epsf

\def\farcm{\hbox{$.\mkern-4mu^\prime$}}
\def\farcs{\hbox{$.\!\!^{\prime\prime}$}}

\begin{document}
\IBVShead{}{May 2005}

\IBVStitletl{Newly Discovered Variable Stars in the Globular Clusters}{NGC~5634, Arp~2 and Terzan~8}

\IBVSauth{SALINAS, R.$^{1,2}$; CATELAN, M.$^1$; SMITH, H.A.$^3$; PRITZL, B.J.$^4$; BORISSOVA, J.$^5$}

\IBVSinst{Pontificia Universidad Cat\'olica de Chile, Departamento de 
          Astronom\'\i a y Astrof\'\i sica, Av. Vicu\~{n}a Mackenna 4860, 
          782-0436 Macul, Santiago, Chile; email: mcatelan@astro.puc.cl}

\IBVSinst{Grupo de Astronom\'\i a, Facultad de Ciencias F\'\i sicas y Matem\'aticas, 
          Universidad de Concepci\'on, Concepci\'on, Chile; email: rcsalina@puc.cl}

\IBVSinst{Department of Physics and Astronomy, Michigan State University, 
          East Lansing, MI 48824, USA; \\ 
          email: smith@pa.msu.edu}

\IBVSinst{Macalester College, 1600 Grand Avenue, Saint Paul, MN 55105, USA; email: pritzl@macalester.edu}

\IBVSinst{European Southern Observatory, Av. Alonso de C\'ordova 3107, 763-0581 Vitacura, Santiago, Chile;
          email: jborisso@eso.org}

\SIMBADobjAlias{NGC~5634, Arp~2, Terzan~8}
\IBVSobj{NGC~5634, Arp~2, Terzan~8}
\IBVSabs{We report the discovery of a sizeable, previously unknown variable star population in the
Sagittarius dwarf spheroidal globular clusters NGC~5634, Arp~2 and Terzan~8. Location, preliminary 
pulsation periods and $B$-band light curves in relative flux units are provided for all these stars.}

\begintext

Cosmological arguments suggest that dwarf galaxies may constitute the fundamental 
building block of larger galaxies. Therefore, comparing the oldest stellar 
components of giant galaxies with those of their surrounding dwarf satellite 
galaxies may provide us with an excellent opportunity to empirically constrain 
the extent to which the dwarfs may have played a role in assembling the giants. 
Unmistakably old and easy to identify in relatively nearby systems, RR Lyrae 
variables may prove of vital importance in this regard (Catelan 2004; Kinman, 
Saha, \& Pier 2004). 
 
In the case of our own Milky Way, several previously unknown dwarf galaxies have 
recently been reported on in the literature, some of which clearly in the process 
of being engulfed by the Milky Way. A particularly striking example is provided 
by the Sagittarius dwarf spheroidal galaxy (Ibata, Lewis, \& Gilmore 1994). An 
important question one must ask is whether the ongoing Sagittarius merger is 
representative of the process that led to the formation of the Milky Way. In 
this sense, comparing the RR Lyrae variables in the globular clusters which 
have been suggested to be associated with Sagittarius (e.g., Da Costa \& 
Armandroff 1995; Dinescu et al. 2000; Bellazzini et al. 2002) and those in the 
Galaxy's halo may shed light on whether the Galactic halo globular cluster system 
originated from Sagittarius-like mergers. In the present note, we focus on four 
Sagittarius-related globulars, namely: Arp~2, Terzan~8, Palomar~12, and NGC~5634.
 
NGC~5634 had seven previously known variables (Baade 1945), while Pal~12 had 
three reported variables (Kinman \& Rosino 1962). Neither Arp~2 nor Terzan~8 have 
variable stars listed in the Clement et al. (2001) catalog. 

Our search for variable stars in these clusters is based on images acquired on 
the Danish 1.54m telescope in La Silla, Chile, over four consecutive nights, 
from June 27 to June 30, 2003. In the course of these nights, the seeing 
conditions varied from 0.9 to 1.5~arcsec. The $2048 \times 2048$ RINGO CCD 
was used. With a pixel scale of $0\farcs 395$, the total observed field was 
$13\farcm 5 \times 13\farcm 5$. 

\vfill\eject
\centerline{Table 1. Locations and tentative periods for new variable stars in NGC~5634.\label{tbl-1}}
\vskip 3mm
\begin{center}
\begin{tabular}{lrrcl}
\hline
Variable  & $x('')$ & $y('')$ & Period (d) & Type   \\ 
\hline
V8       & $55.3$   & $-43.4$ & 0.330     & RRc  \\ 
V9       & $0.4$    & $-4.7$  & 0.583     & RRab \\  
V10       & $13.0$   & $-0.8$  & 0.646     & RRab \\
V11       & $10.7$   & $7.9$   & 0.660     & RRab \\
V12       & $-2.4$   & $9.5$   & 0.624     & RRab \\
V13       & $-14.2$  & $12.6$  & 0.645     & RRab \\
V14       & $21.3$   & $18.6$  & 0.720     & RRab \\
V15       & $7.9$    & $18.6$  & 0.852     & RRab \\ 
V16       & $9.9$    & $31.6$  & 0.670     & RRab \\  
V17      & $-0.4$   & $43.1$  & 0.289     & RRc  \\
V18      & $-20.9$  & $-38.3$ & 0.325     & RRc  \\
V19      & $0.8$    & $-26.1$ & 0.296     & RRc  \\
V20      & $-7.9$   & $0.0$   & 0.648     & RRab \\
V21      & $-30.8$  & $-28.0$ & 0.0666     & SX~Phe\\
  
\hline
\end{tabular}
\end{center}

\vskip 0.5cm

\centerline{Table 2. Locations and tentative periods for new variable stars in Arp~2.\label{tbl-2}}
\vskip 3mm
\begin{center}
\begin{tabular}{lrrcll}
\hline
Variable  & $x('')$ & $y('')$ & Period (d) & Type  & Note\\
\hline
V1       & $-101.1$ & $27.6$  & 0.568     & RRab & Valenti's V4\\
V2       & $-58.1$  & $73.5$  & 0.821     & RRab & Valenti's V5\\
V3       & $160.4$  & $-27.6$ & 0.565     & RRab & Valenti's V25\\
V4       & $223.2$  & $1.2$   & 0.458     & RRab & Valenti's V28\\
V5       & $128.8$  & $-327.5$& 0.763     & RRab \\
V6       & $-190.8$ & $-100.7$& 0.445     & RRab \\
V7       & $289.1$  & $288.0$ & 0.530     & RRab \\
V8       & $4.0$    & $-40.0$ & 0.292     & RRc  \\
V9       & $97.6$   & $-63.6$ & 0.517     & RRab \\
V10      & $90.1$   & $-125.2$& 0.0473     & SX~Phe\\
V11      & $20.9$   & $68.3$  & 0.0611     & SX~Phe\\
V12      & $-43.8$  & $-237.0$& 0.0604     & SX~Phe \\

\hline
\end{tabular}
\end{center}

\vskip 0.5cm

\centerline{Table 3. Locations and tentative periods for new variable stars in Terzan~8.\label{tbl-3}}
\vskip 3mm
\begin{center}
\begin{tabular}{lrrcll}
\hline
Variable  & $x('')$ & $y('')$ & Period (d) & Type   & Note\\
\hline
V1       & $-113.4$ & $-187.6$  & 0.686  & RRab   & Montegriffo's 117\\
V2       & $124.0$  & $-23.3$  & 0.392   & RRc    & Montegriffo's 1350\\
V3       & $-179.0$  & $-193.2$  & 0.601   & RRab\\
V4       & $95.6$   & $37.1$  & 0.0616      & SX~Phe    \\

\hline
\end{tabular}
\end{center}

\vskip 1.0cm

The total set of images consists in 32 $B$, $V$ pairs for NGC~5634, 37 pairs for 
Arp~2, 27 pairs for Ter~8, and 34 pairs for Pal~12. In this note we shall restrict  
ourselves to relative-flux light curves based on the $B$-band images.

Using the image subtraction technique (ISIS v2.1; Alard 2000), we were able to 
re-discover six of the seven known variables in NGC~5634, and to discover 14 new 
variables in the cluster. In Arp~2, we discovered 8 new variables and confirmed 
4 previously reported ones (Valenti 2001). Other variables reported by Valenti  
were not found to be variable in our data. In Ter~8 two new variables 
were found and a two more previously suspected variables (Montegriffo et al. 1998) 
were confirmed. We do not confirm the variable status of stars V1, V2, and V3 
that had previously been reported in Pal~12 (Kinman \& Rosino 1962). 

The location, classification and tentative periods for the new variables in 
NGC~5634, Arp~2, and Ter~8 are given in Tables~1,~2, and 3, respectively. 
In these tables, the $x$ and $y$ coordinates are in arcseconds with respect to 
the cluster centers, as given in the online Clement et al. (2001) catalog. 
Because the time coverage was not extensive, the periods are probably good only 
to the third decimal place, and some may actually be aliases of the correct 
period. Light curves based on the reported periods are shown in Figure~1 
for NGC~5634, and in Figure~2 for Arp~2 (first three rows) and Ter~8 (bottom row). 
Finding charts for the newly discovered variable stars in the three clusters are 
provided in Figures~3 and 4 (NGC~5634), Figure~5 (Arp~2), and Figure~6 (Ter~8). 

One of the variables found by Baade (1945) in NGC~5634, V7, could not be confirmed  
by our analysis. Likewise, the variable was not present in the Liller \& Sawyer-Hogg 
(1976) analysis, although in their case this was due to blending. Since Alard's 
(2000) image subtraction technique is particulary powerful in the center of 
concentrated clusters, we conclude that V7 is not a variable. For V2 in NGC~5634, 
we find that the period given by Liller \& Sawyer-Hogg ($P=0.605148$~d) 
does not give us a clean light curve; $P=0.601$~d provides a better solution. 
For the remainder of the variables, the periods that we found are the same as in 
Liller \& Sawyer-Hogg's study. Of the twelve variables discovered in Arp~2, four 
(NV1-NV4) had previously been found by Valenti (2001). Two of the Ter~8 variables 
(NV1-NV2) were previously identified by Montegriffo et al. (1998). Since their studies 
were based on very few datapoints, our periods represent a significant improvement 
over the ones previously reported. 

Note that we detect candidate SX~Phoenicis variables in NGC~5634, Arp~2, and 
Ter~8. Although their light curves tend to be a bit noisy, they are all located 
in the blue straggler region in their respective clusters' color-magnitude diagrams 
(Salinas et al. 2005, in preparation); together with their short periods, this 
suggests to us that their SX~Phe clasification is reliable. 

In a future paper, we will attempt to incorporate additional data into our analysis, 
calibrate the light curves into standard magnitudes, construct Bailey diagrams and 
analyze the Sagittarius globular cluster system in the context of the Oosterhoff 
dichotomy and the formation history of the Milky Way (Catelan 2004, 2005).

\IBVSTack{We thank C. Cacciari and E. Valenti for interesting discussions. 
R.S. acknowledges support by FONDAP 15010003. M.C. acknowledges support by Proyecto 
FONDECYT Regular No.~1030954. H.A.S. 
acknowledges the NSF for support under grant AST 02-05813.}


\references

Alard, C., 2000, {\it A\&AS}, {\bf 144}, 235 

Baade, W., 1945, {\it ApJ}, {\bf 102}, 17


Bellazzini, M., Ferraro, F.R., Ibata, R. 2002, {\it AJ},
{\bf 124}, 915

Catelan, M., 2004, in {\it Variable Stars in the Local Group}, IAU Colloq. 193, 
  ed. D.W. Kurtz \& K.R. Pollard, ASP Conf. Ser., {\bf 310}, 113
  (San Francisco: ASP)

Catelan, M., 2005, in {\it Resolved Stellar Populations},  
  ed. D. Valls-Gabaud \& M. Ch\'avez, ASP Conf. Ser., in press (astro-ph/0507464)

Clement, C.M., et al., 2001, {\it AJ}, {\bf 122}, 2587

Da Costa, G.S., Armandroff, T.E., 1995, {\it AJ}, {\bf 109}, 2533

Dinescu, D., Majewski, S., Girard, T., Cudworth, K., 
 2000, {\it AJ}, {\bf120}, 1892

Ibata, R., Gilmore, G., Irwin, M., 1994, {\it Nature}, {\bf 370}, 194

Kinman, T.D., Rosino, L.\ 1962, {\it PASP}, {\bf 74}, 499 

Kinman, T.D., Saha, A., Pier, J.R., 2004, {\it ApJ}, {\bf 605}, L25

Liller, M., Sawyer Hogg, H., 1976, {\it AJ}, {\bf 81},628

Montegriffo, P., Bellazzini, M., Ferraro, F.R., et al., 1998, {\it MNRAS}, {\bf 294}, 315

Valenti, E., 2001, {\it Ricerca delle variabili RR Lyrae
neglie amassi globulari NGC 6304 e Arp~2}, Tesi di Laurea
(Universit\`{a} degli studi di Bologna, Bologna)

\endreferences

\IBVSfig{6.5in}{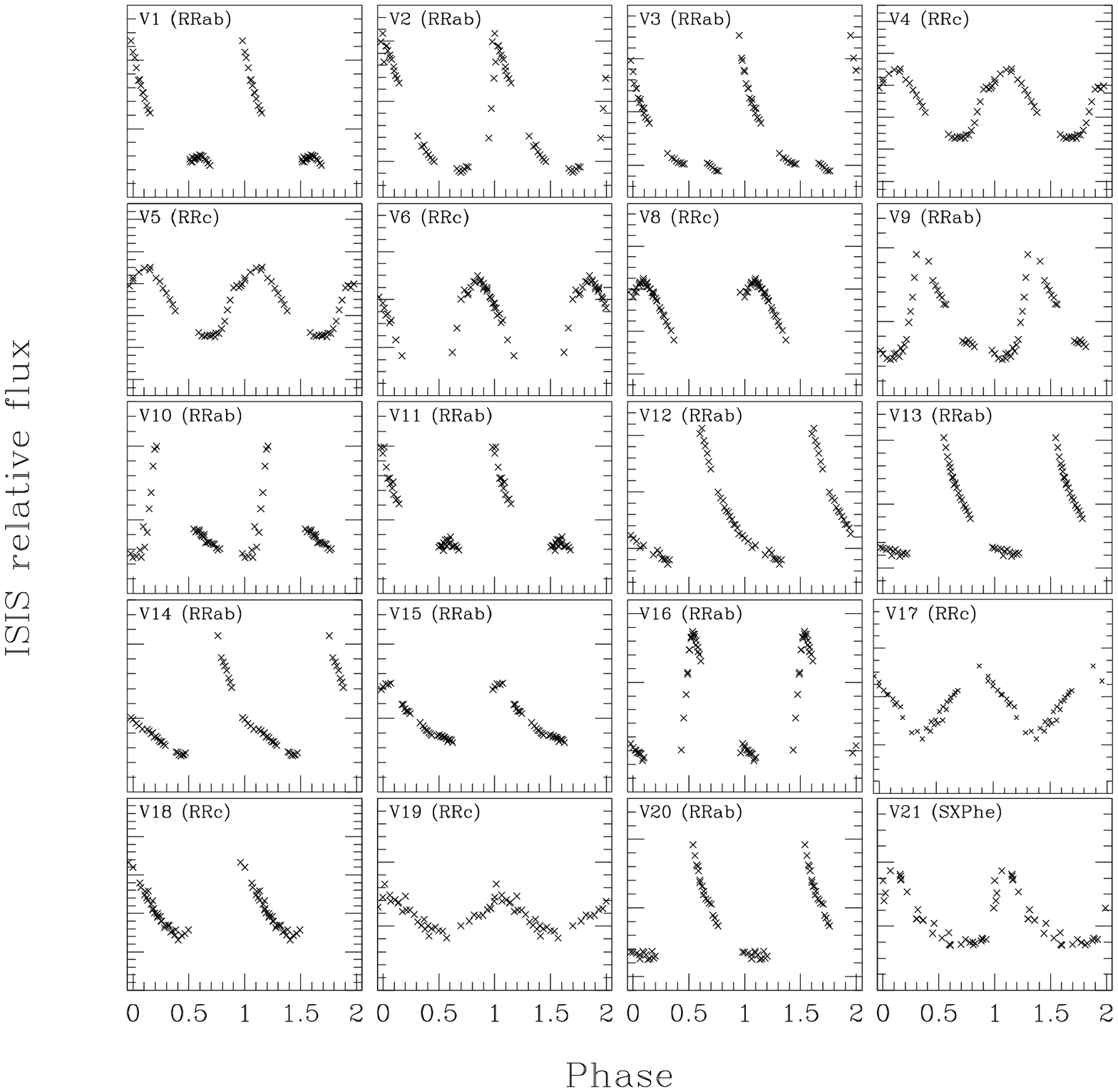}{$B$-band differential light curves for six previously known RR Lyrae variables and fourteen newly discovered variables in NGC~5634.}

\IBVSfig{6.5in}{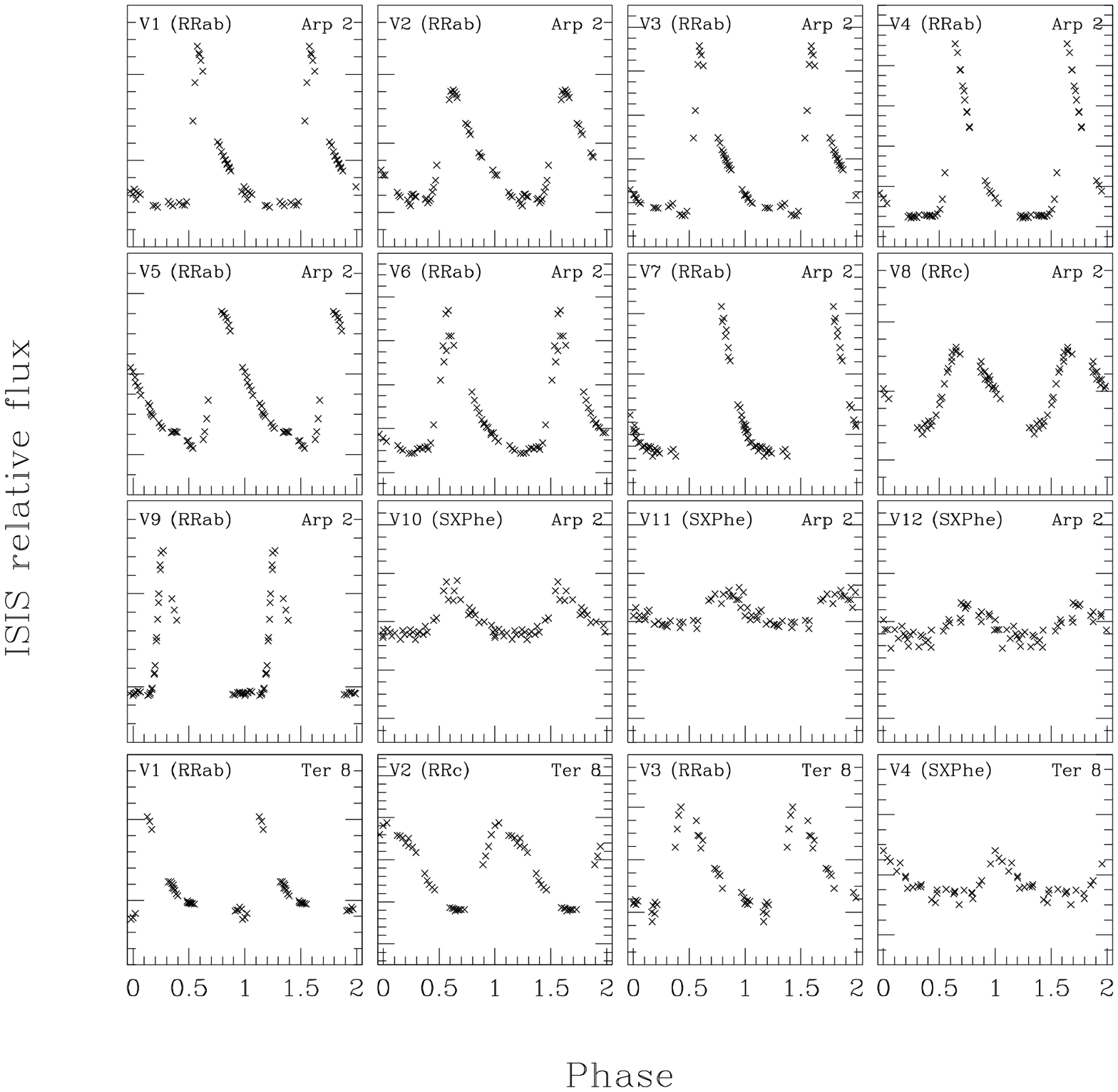}{$B$-band differential light curves for the twelve variables detected in Arp~2 and four variables in Terzan 8 (last row).}

\IBVSfig{5in}{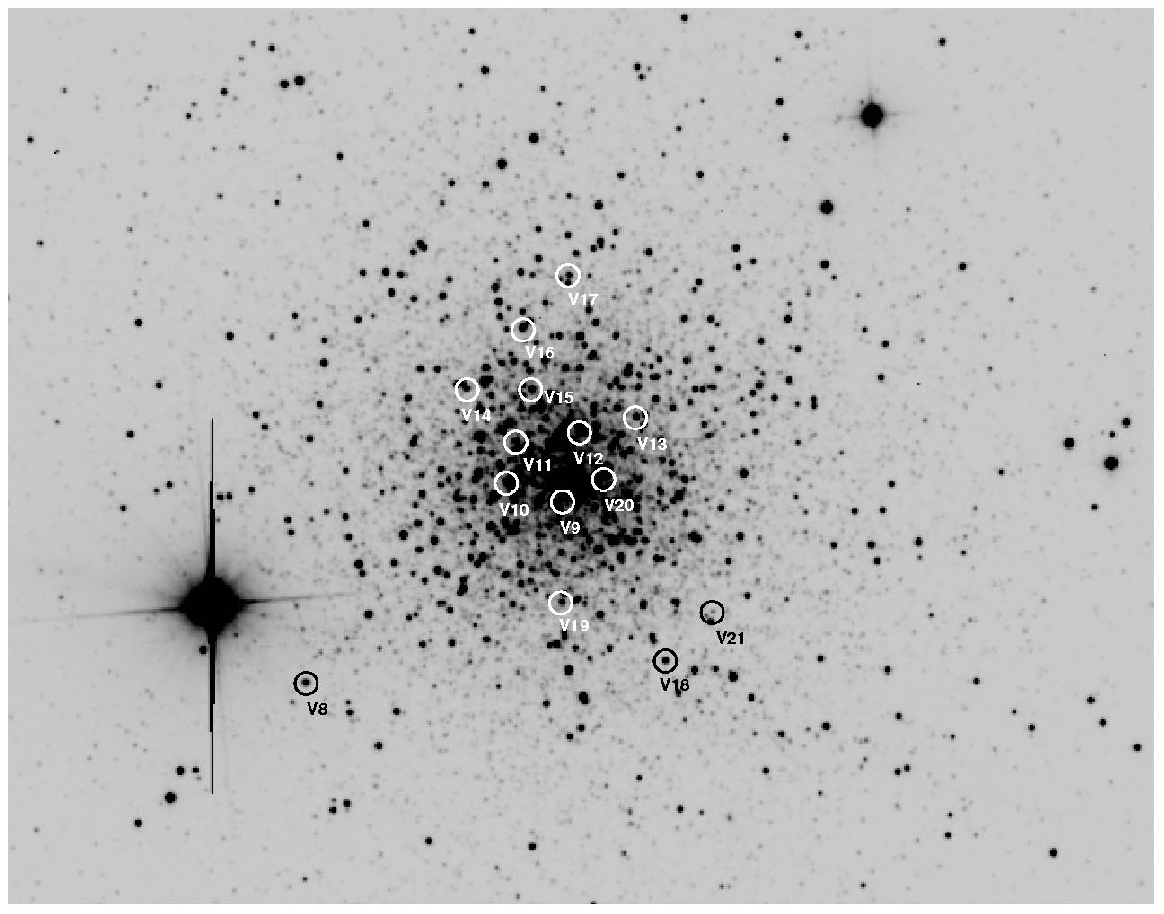}{Finding chart for newly discovered variable stars in NGC~5634. 
   The field size is approximately $4'\times 3'$. North is up and East to the left.}

\IBVSfig{5in}{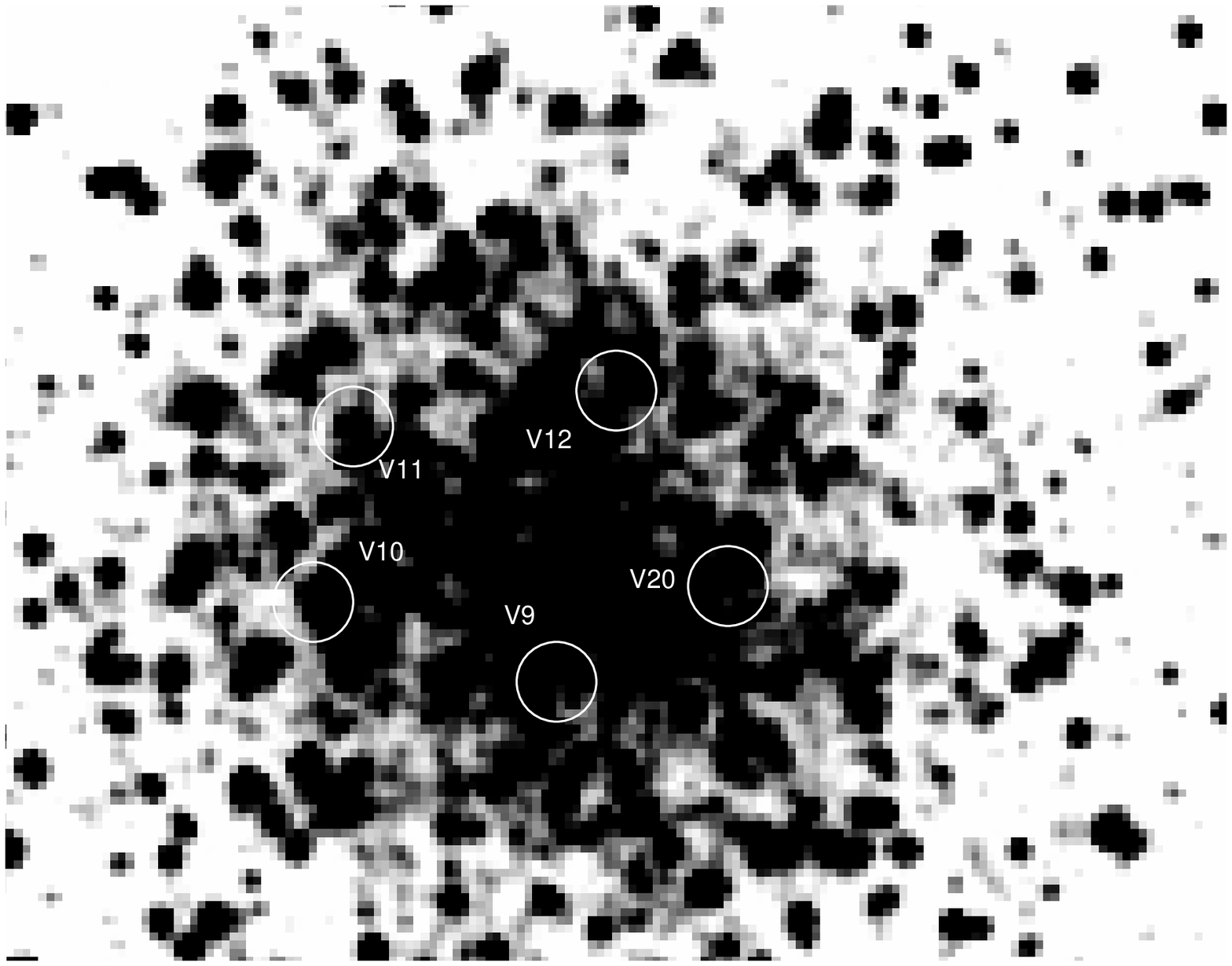}{Same as in the previous plot, but zooming in on a central 
region of approximately $1'\times 1'$ in size. This plot illustrates the power of ISIS
to reliably detect and perform relative photometry for variable stars in extremely
crowded fields.}

\IBVSfig{5in}{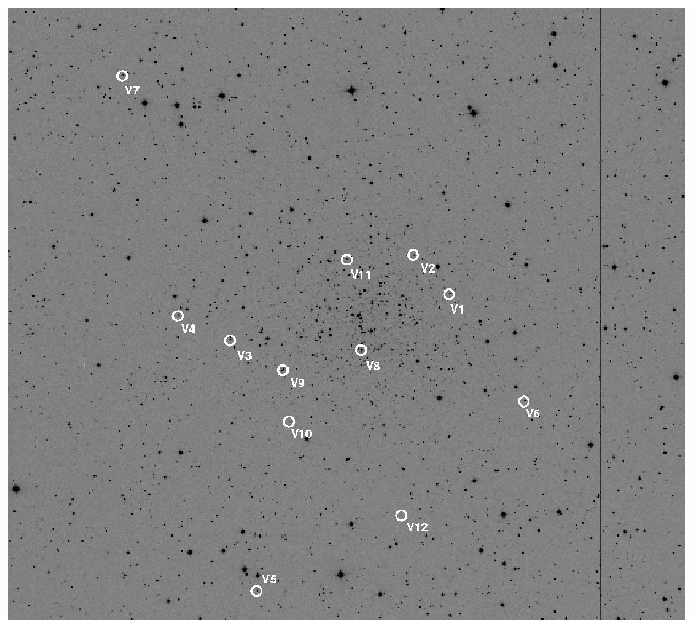}{Finding chart for newly discovered variable stars in Arp~2.
   The field size is approximately $13'\times 13'$. North is up and East to the left.}

\IBVSfig{5in}{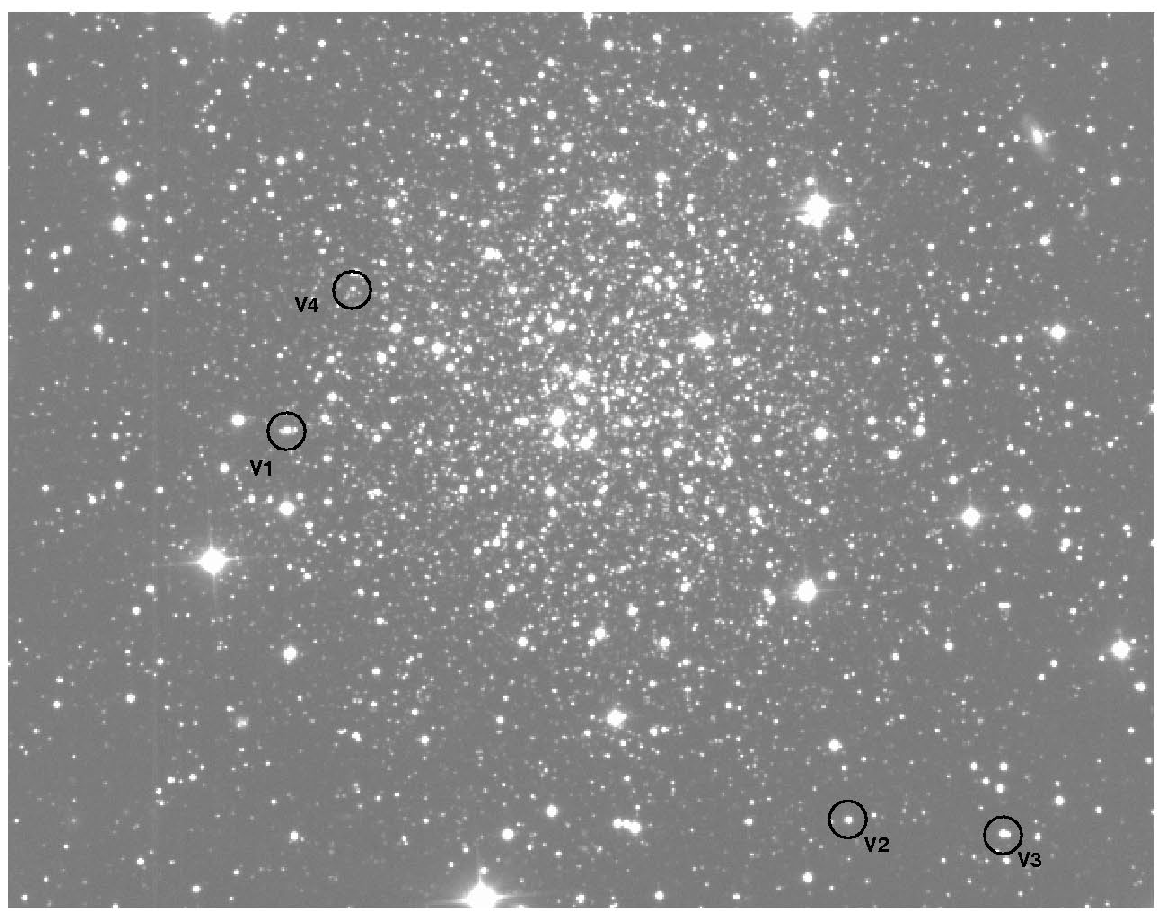}{Finding chart for newly discovered variables in Terzan~8.
   The field size is approximately $8'\times 6'$. North is up and East to the left.}

\end{document}